\pdfoutput=1
\documentclass[nofootinbib]{revtex4}
\usepackage{latexsym}
\usepackage{amsfonts}
\usepackage[latin1]{inputenc}
\usepackage[caption=false]{subfig}
\usepackage{graphicx}
\usepackage{mathrsfs}
\usepackage{amssymb}
\usepackage{amsmath}
\usepackage{color}
\usepackage{fancyhdr}
\usepackage{color}

\begin{document}
\begin{flushright}
PSI-PR-13-13
\end{flushright}

\title{Higgs Boson in the 4DCHM: LHC phenomenology}

%

\author{D. Barducci, A. Belyaev, M.S. Brown, S. Moretti}
\affiliation{School of Physics and Astronomy, University of Southampton, Highfield, SO17 1BJ, UK.}
\email{d.barducci,a.belyaev,m.s.brown,s.moretti@soton.ac.uk}
\author{S. De Curtis}
\affiliation{INFN, Sezione di Firenze,
Via G. Sansone 1, 50019 Sesto Fiorentino, Italy.}
\email{decurtis@fi.infn.it}
\author{G.M. Pruna}
\affiliation{Paul Scherrer Institute, CH-5232 Villigen PSI, Switzerland.}
\email{giovanni-marco.pruna@psi.ch}

\begin{abstract}
\noindent
Composite Higgs models provide an elegant solution to the hierarchy problem present in the Standard Model (SM)
and give an alternative pattern leading to the mechanism of Electro-Weak Symmetry Breaking (EWSB).
We present an analysis of the Higgs boson production and decay within a recently proposed realistic realization of this general idea: the 4D Composite Higgs Model (4DCHM).
Comparing our results with the latest Large Hadron Collider (LHC) data we show that the 4DCHM
could provide an alternative explanation with respect to the SM of the LHC results pointing to the discovery of a Higgs-like particle at 125 GeV.
\end{abstract}

\maketitle

\thispagestyle{fancy}


\section{The Higgs as a Pseudo Nambu-Goldstone Boson (PNGB) and the 4DCHM}
After the recent discovery of a Higgs-like state at the LHC one of the primary question to ask is whether these new boson is compatible with the one predicted in the SM or if it belongs to some other Beyond the SM (BSM) theory.
BSM scenarios are usually asked to solve the (in)famous hierarchy problem present in the SM and the most common way to do this is to add a new kind of symmetry.
Besides supersymmetry another solution based on the Goldstone symmetry has been more and more taken into account in the last years: this is the idea at the basis of composite Higgs models in which the Higgs boson is a PNGB arising from the spontaneous breaking of a global symmetry.
Modified Higgs couplings with respect to the SM ones and the presence of extra gauge bosons and coloured fermions can affect the Higgs phenomenology
making these scenarios really attractive in the light of the LHC results.

A 4 dimensional realization of the general idea proposed in \cite{Kaplan:1983fs} and specialized to the coset $SO(5)\rightarrow SO(4)$ in \cite{Agashe:2004rs} has been proposed in \cite{DeCurtis:2011yx} under the name the 4DCHM name and will be the framework of
our analysis \cite{Barducci:2013wjc,Barducci:2013ica}.

Besides the SM particles the 4DCHM present in its spectrum 8 extra gauge bosons, 5 neutral called $Z^\prime$ and 3 charged called $W^\prime$, and 20 new quarks: 8 with charge $+2/3$, 8 with charge $-1/3$ and 2 respectively with charge $5/3$ and $-4/3$; these coloured fermions are called $t^\prime$, $b^\prime$,
$\tilde{T}$ and $\tilde{B}$, respectively.
These states mix via the \emph{partial compositness}  mechanism with the SM states having the same $SU(3)_c \otimes U(1)_{em}$ charge and this together with the non linear realization of the Goldstone symmetry give rise to Higgs couplings which are modified with respect to the SM. Moreover in loop induced processes the effective couplings are also modified by the presence of extra states running into the loops (as in the $H\rightarrow gg$ and $H\rightarrow \gamma\gamma$ vertices for example).

\section{$\kappa$ parameters, $\mu$ parameters and fit to experimental data}

The parameter space of the model is given in terms of the scale $f$ of the spontaneous symmetry breaking $SO(5)\rightarrow SO(4)$, the $g_*$ coupling of the new $SO(5)$ gauge fields and a set of 9 parameters in the fermionic sector which are a mass parameter of the extra states, $m_*$, 4 mixing parameters between SM and extra quarks, $\Delta^{T,B}_{L,R}$ and 4 Yukawa parameters describing the interaction between the extra states themselves, $Y_{T,B}$ and $m_{Y_{T,B}}$.
Given the large number of parameters and particles the model has been implemented in automatic tools (using LanHEP \cite{Semenov:2010qt} and CalcHEP \cite{Belyaev:2012qa}).
The parameter scan has been done for various model scales $f$, fixing $m_{Z^\prime,W^\prime}\simeq2$ TeV, varying the parameter of the fermionic sector and imposing to reconstruct the Higgs, top and bottom masses. LHC limits arising from the direct search of extra quarks have been taken into account.

To compare our results to LHC data we have defined the following parameters, the latter being the signal strength that is the observed signal over the SM expectation, recasted in term of the $\kappa$ parameters\footnote{$Y^\prime$ denotes the particles partecipating in the Higgs boson production, e.g. $gg$ in the case of gluon fusion.}:
\begin{equation}
\kappa^2_{b,g,V,\gamma}=\frac{\Gamma(H\rightarrow b \bar b, gg, VV, \gamma\gamma)_{{\scriptsize{\textrm{4DCHM}}}}}{\Gamma(H\rightarrow b \bar b, gg, VV, \gamma\gamma)_{{\scriptsize{\textrm{SM}}}}}
\quad \kappa^2_H = \frac{\Gamma_{tot}(H)_{{\scriptsize{\textrm{4DCHM}}}}}{\Gamma_{tot}(H)_{{\scriptsize{\textrm{SM}}}}}
\quad \mu^{Y^\prime Y^\prime}_{YY}=\frac{\kappa^2_{Y^\prime}\kappa^2_Y}{\kappa^2_H}
\label{c}
\end{equation}

\begin{figure}[h!]
 \begin{center}
\includegraphics[width=4.1cm]{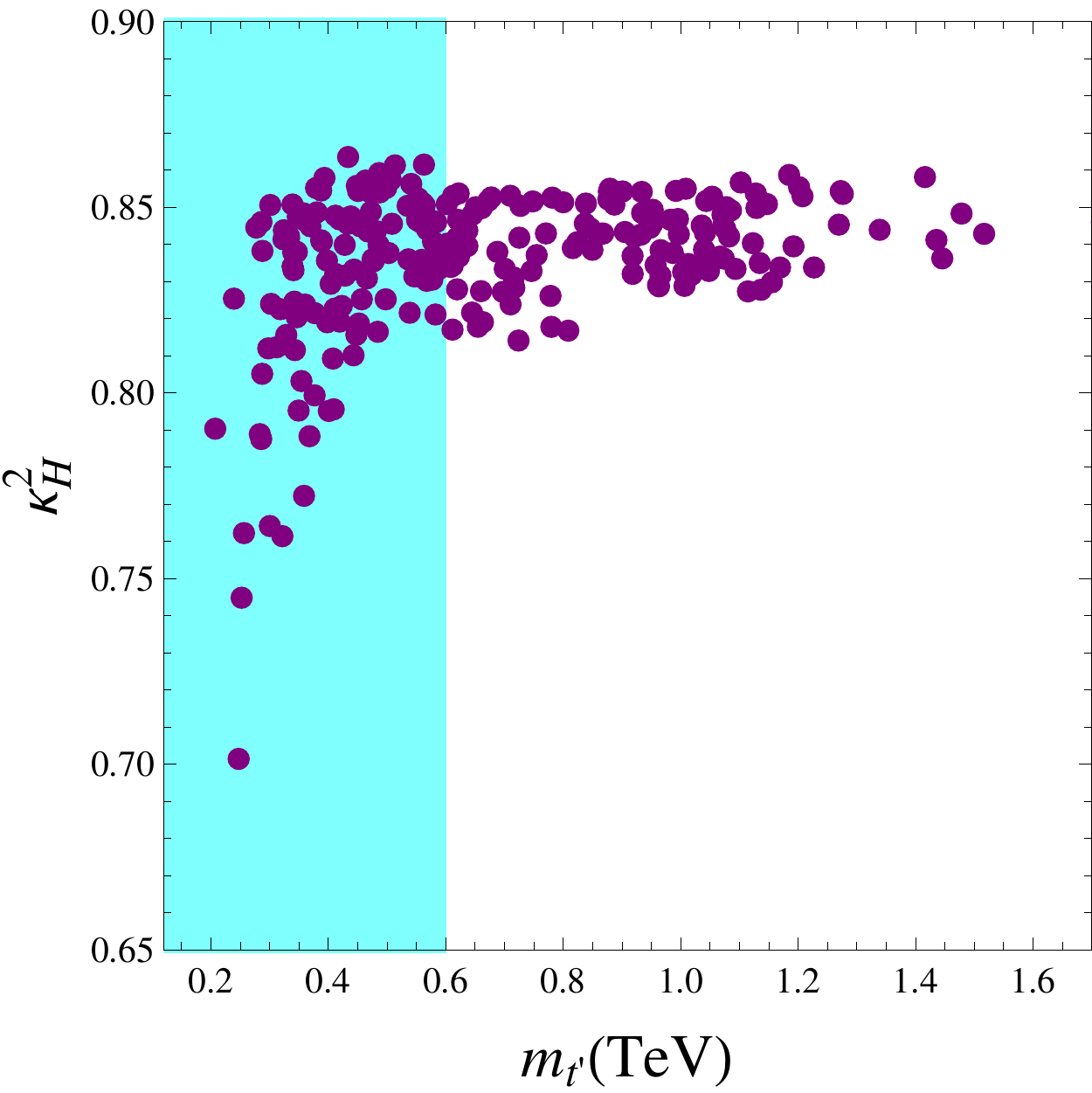}
\includegraphics[width=4.1cm]{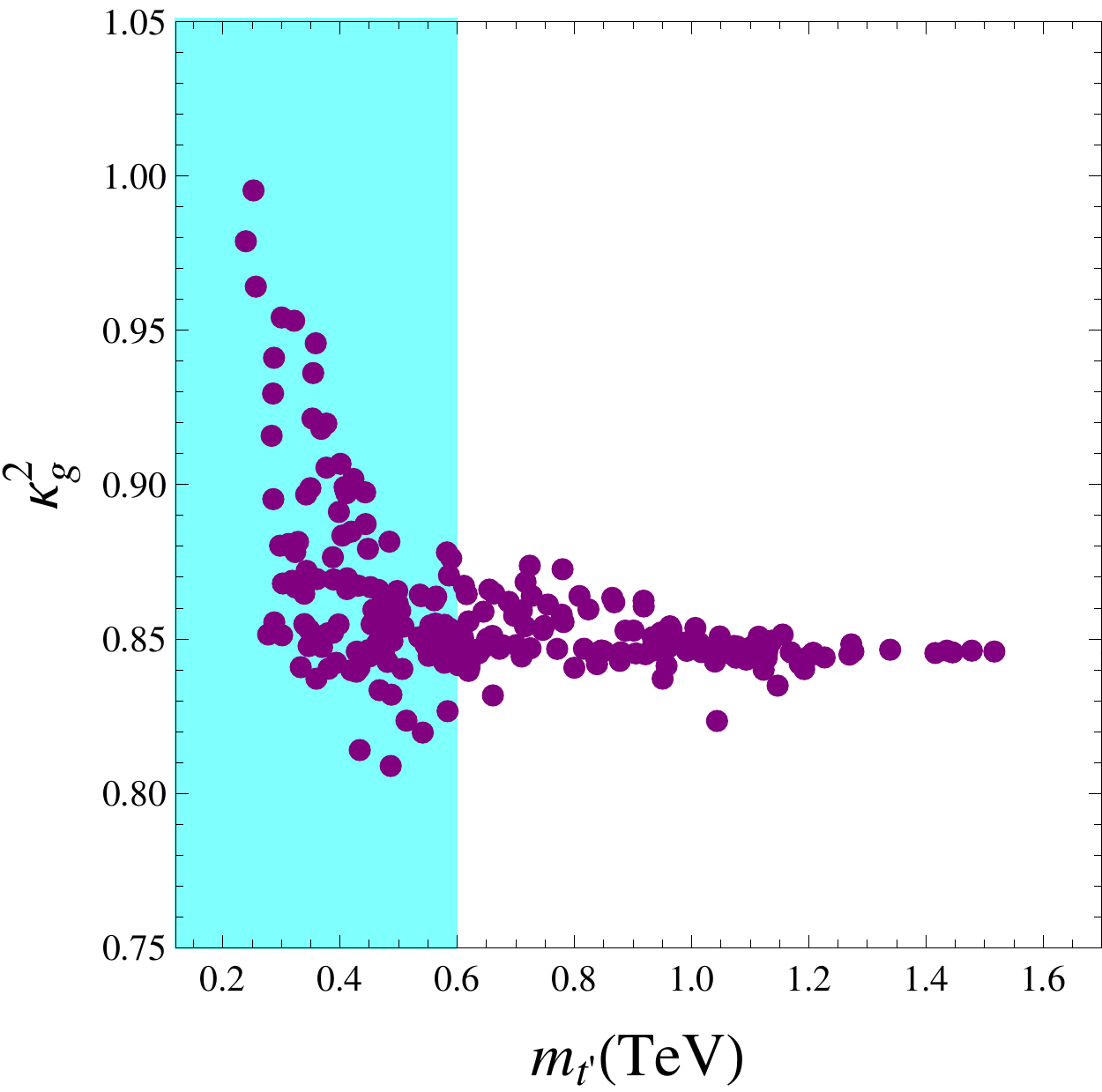}
\includegraphics[width=4cm]{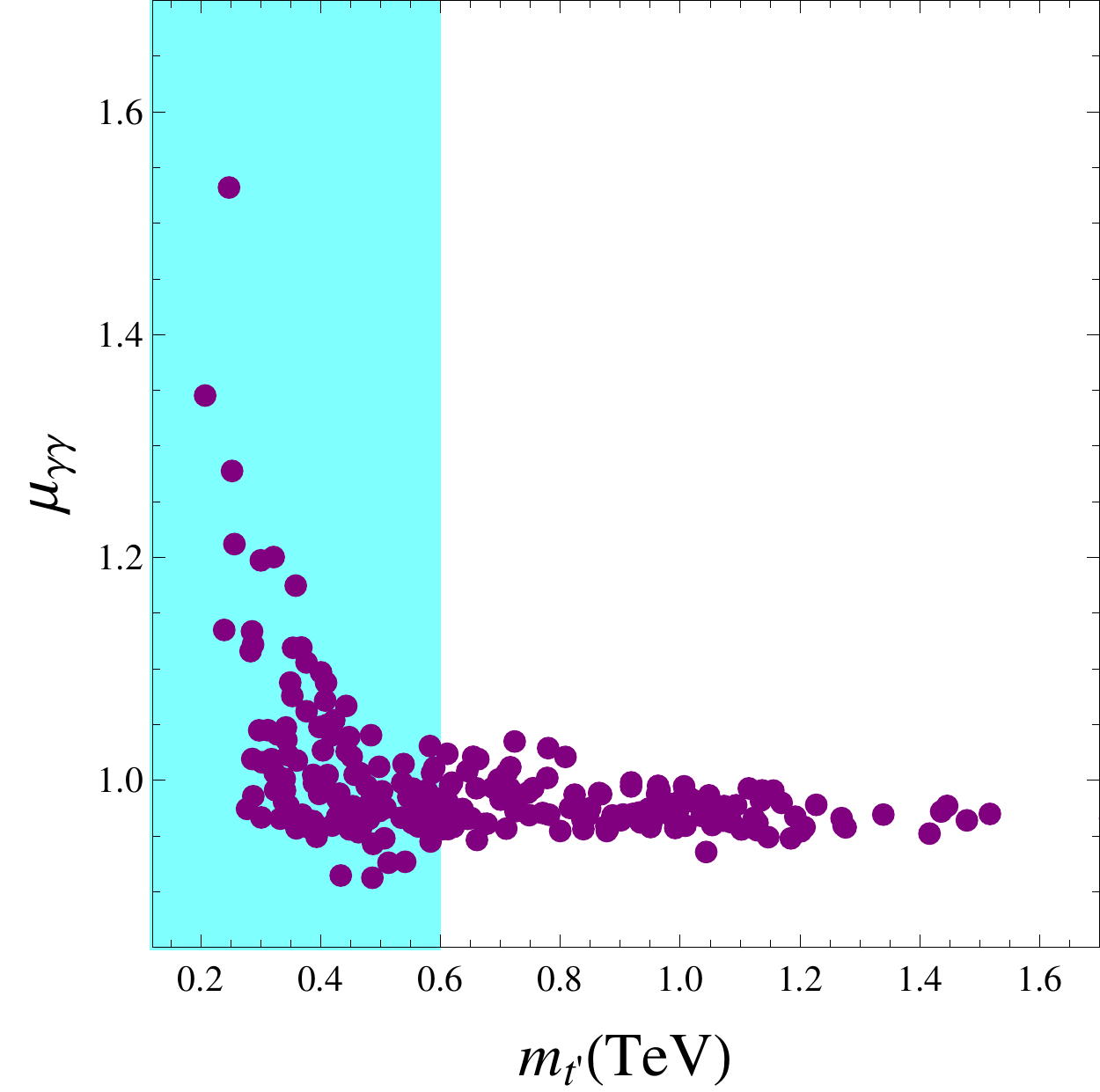}
\end{center}
\caption{Distributions of $\kappa^2_H$ (left), $\kappa^2_{gg}$ (centre) and $\mu_{\gamma\gamma}$ (right) as a function of the lightest 2/3 fermion resonance for the choice $f=1$ TeV. The shaded region corresponds to masses of the fermionic partner excluded by LHC direct searches. }
\label{fig:kmupar}
\end{figure}

In Fig.\ref{fig:kmupar} we show the plots for $\kappa^2_H$ (left), $\kappa^2_{gg}$ (centre) and $\mu_{\gamma\gamma}$ (right) where one notices the decrease of the Higgs total width and $gg$ partial width with respect to the SM mainly due to the modified $g_{H b\bar b/H t \bar t}$ coupling. $\kappa^2_H $ also enters the computation of the $\mu_{\gamma\gamma}$ parameter compensating the decrease of the effective $g_{Hgg}$ and $g_{H\gamma\gamma}$ couplings and giving a total result almost stable with respect to the SM.

We have then performed a $\chi^2$ fit for various choices of the model scale $f$ with respect to the signal strength parameters given by the ATLAS and CMS collaborations.
Our results are found in Fig. 2.

 \begin{figure}[h!]
 \begin{center}
 \includegraphics[width=3.6cm]{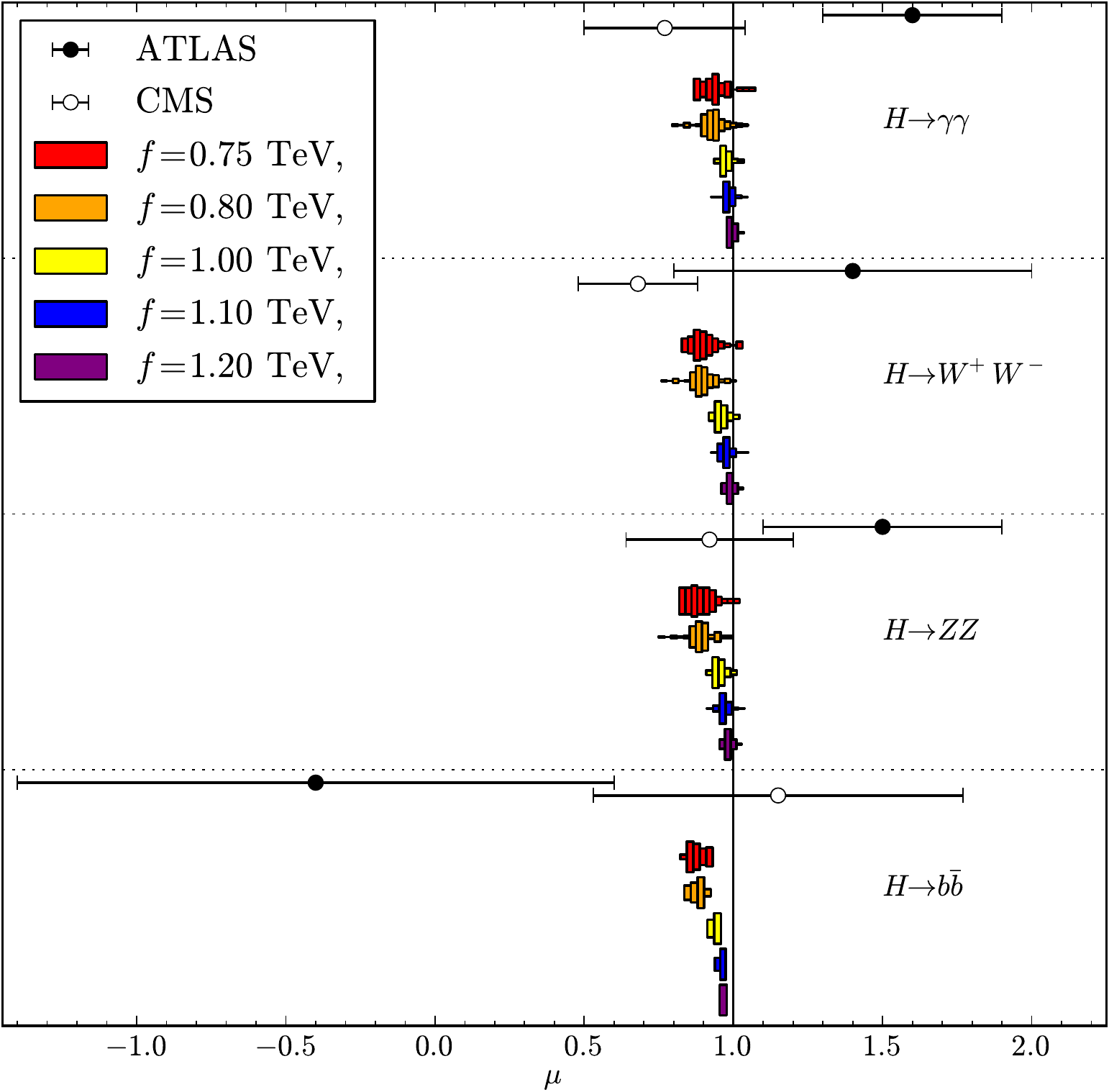}
\includegraphics[width=4.608cm]{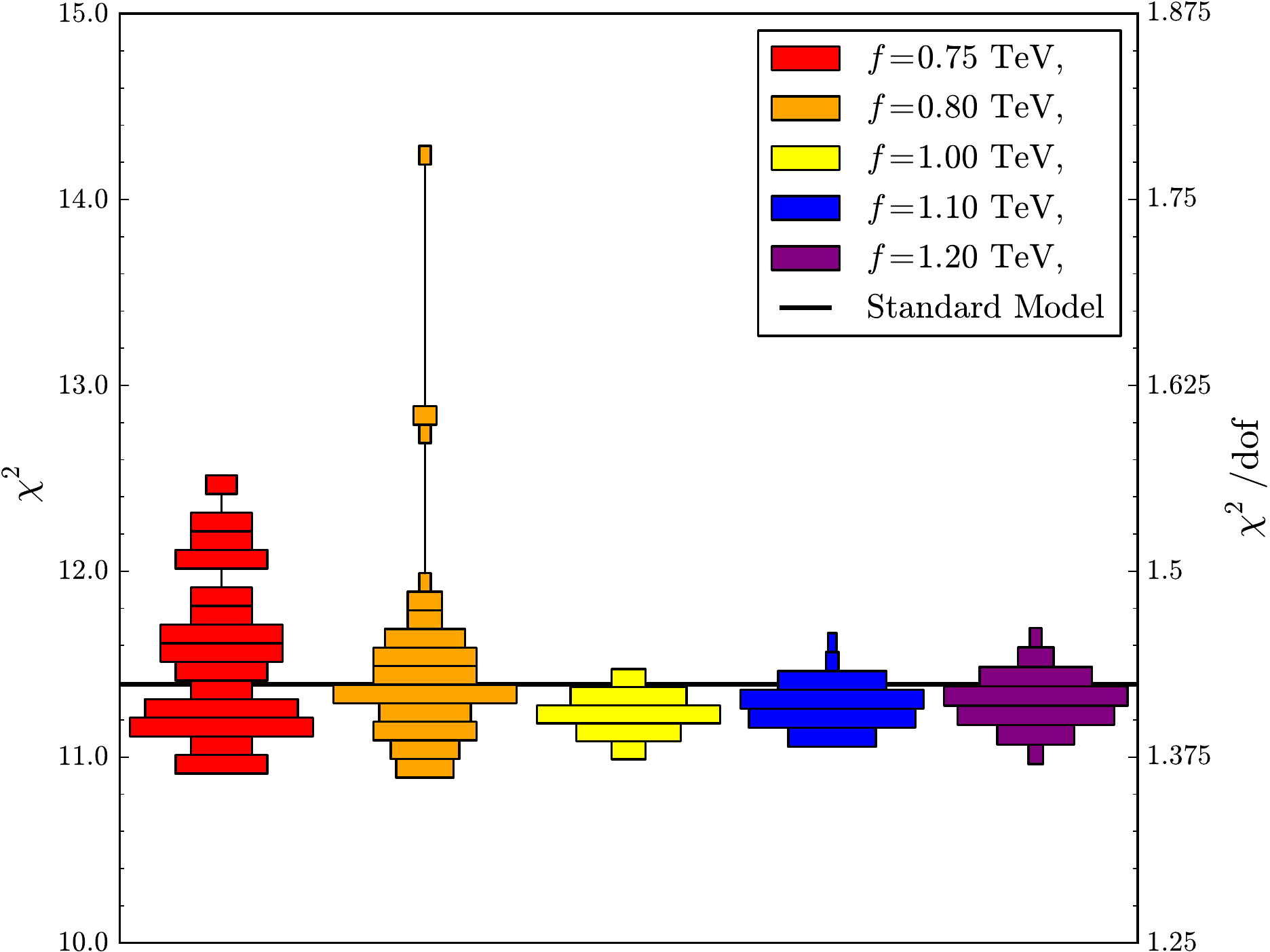}
\end{center}
\caption{Comparison of the $\mu$ parameters of eq. (2.1) with the experimental values by ATLAS and CMS (left) and the $\chi^2$ fit to this data where the horizontal black line represents the SM (right) for various choices of the model scale $f$. All points here are compliant with the LHC direct search limits on extra quarks.}
\label{fig:chifit}
\end{figure}

\newpage

\section{Conclusions}
In this proceeding we presented an analysis of the scalar sector of the 4DCHM and illustrated how the Higgs couplings and the signal strengths are modified with respect to the SM. Further, after having performed a $\chi^2$ fit, we described how the 4DCHM could provide an alternative explanation to the LHC data pointing to a Higgs boson discovery.


\end{document}